\documentclass[aps,pre,reprint,twocolumn,noshowpacs,floatfix,longbibliography]{revtex4-1}
\usepackage[T1]{fontenc}
\usepackage[latin9]{inputenc}
\usepackage{siunitx}
\usepackage{float}
\usepackage{amsmath}
\usepackage{graphicx}
\usepackage[dvipsnames]{xcolor}

\begin{document}

\title{Local rules for fabricating allosteric networks}

\author{Nidhi Pashine} 
\email{npashine@uchicago.edu}
\affiliation{Department of Physics, The University of Chicago, Chicago IL, 60637, USA}

\begin{abstract}

Mechanical properties of disordered networks can be significantly tailored by modifying a small fraction of their bonds. This procedure has been used to design and build mechanical metamaterials with a variety of responses. A long-range `allosteric' response, where a localized input strain at one site gives rise to a localized output strain at a distant site, has been of particular interest. This work presents a novel approach to incorporating allosteric responses in experimental systems by pruning disordered networks \textit{in-situ}. Previous work has relied on computer simulations to design and predict the response of such systems using a cost function where the response of the entire network to each bond removal is used at each step to determine which bond to prune. It is not feasible to follow such a design protocol in experiments where one has access only to local response at each site. This paper presents design algorithms that allow determination of what bonds to prune based purely on the local stresses in the network without employing a cost function; using only local information, allosteric networks are designed in simulations and then built out of real materials. The results show that some pruning strategies work better than others when translated into an experimental system. A method is presented to measure local stresses experimentally in disordered networks. This approach is then used to implement pruning methods to design desired responses \textit{in-situ}. Results from these experiments confirm that the pruning methods are robust and work in a real laboratory material.

\end{abstract}
\maketitle

\section{Introduction}
Recent advances in the field of mechanical metamaterials have shown that disordered networks are extremely tunable so that their mechanical response can be altered dramatically by modifying a small number of edges or bonds between nodes. This can be demonstrated in the Poisson's ratio, $\nu$, which is the negative of the ratio of the strain along the transverse axes to an applied strain along a given axis.  In an isotropic network material in $d$ dimensions, $\nu$ can be varied between the two theoretical limits, $\nu=-1$ (auxetic) and $\nu= +1/(d-1)$ (incompressible), by selectively removing a small fraction of the network bonds~\cite{goodrich2015principle,hexner2018a,hexner2018b}. 

A more general property that can be incorporated into a disordered network is a long-distance response, where applying an input strain at a local site in the system creates an output strain at another distant localized site~\cite{rocks2017designing, yan2017architecture,tlusty2017}. This is referred to as a mechanical `allosteric' response because it is inspired by the property of allostery in protein molecules. 

Both allosteric and auxetic responses have been successfully designed and incorporated into physical networks by pruning selected bonds~\cite{rocks2017designing, reid2018auxetic}. An important difference between auxetic and allosteric response is that the Poisson's ratio is a monotonic function of the ratio of the shear, $G$, and bulk, $B$, moduli of a material. In a disordered system, once contributions of every bond to the bulk and shear moduli are known, it is straightforward to change its Poisson's ratio by pruning specific bonds. 

On the other hand, in earlier works, allosteric systems have been designed by removing bonds from a disordered network using a cost function. This protocol succeeds well for designing materials with multiple targets controlled by a single source using computer simulations~\cite{rocks2017designing,rocks2019}. A cost function calculates the global response of the system due to the removal of each bond individually in the network in order to decide what bonds need to be pruned to minimize the difference from a desired response. Such a cost function is difficult to interpret or quantify in terms of simple local properties of the individual bonds in the network. This makes it difficult to create such behavior in a network \textit{in situ} so that the result can be achieved without recourse to prior design on a computer.  

This paper takes an alternate approach for designing a pruning protocol; the aim is to use only local information encoded in the stresses on each bond due to an externally applied strain.
%As in the case of auxetic response, %the goal is to determine how each bond in a network contributes to the desired interaction. T
This would allow the creation of allosteric responses in spring-network simulations by using only local information before a bond is removed.  This approach is a generalization of the one used to incorporate auxetic response into networks~\cite{goodrich2015principle}. In that case, the pruning was based on the local stresses in the bonds due to an externally applied strain.  In the case of allostery, the procedure is extended to include the response to a set of separate, individually applied, strains which are then combined.   The results are then tested and validated in experiments; I take the networks that were designed in simulations and build them out of rubber sheets. 

One problem encountered in using simulated networks to prune real materials is that the simulations used, which have been of disordered central-spring networks derived from jammed packings of spheres~\cite{liu2010jamming}, are overly simplified models of the real materials.   
%and incorporate allosteric responses into these systems using various pruning methods.
A physical system is more complicated than such a spring network because it has forces other than those derived from harmonic central-spring interactions. To circumvent this problem, I present an experimental approach to measure the relative magnitude of stresses in networks under any external strain and use it to prune the network systems \textit{in-situ}. This is done using photoelastic networks that are observed between pairs of cross-polarizers.  In this approach, no simulations are necessary for determining which bonds to prune.

I find that experimental networks that are designed in simulations have a drastically different response than ones that are pruned \textit{in-situ} using photoelastic stress measurements.  In addition, the different local pruning methods can produce different results in experiments even when they are designed to give the same response in simulations. 
Taken together, this work improves the understanding of the mechanisms that control allostery in mechanical systems and opens up possibilities of building new and interesting mechanical responses in real materials.

\section{Theoretical Approach}
The random disordered spring networks are created in two dimensions, $2D$, with periodic boundary conditions. These networks are derived from $2D$ jammed packings of soft discs which are under force balance~\cite{liu2010jamming,vanHecke_2015,Ohern}. Each point of contact between the discs is replaced by a harmonic spring that connects the centers of the two discs. The equilibrium length of each spring is chosen to be the distance between the centers of the discs. This ensures that the resultant network of nodes connected by bonds is under zero stress in its ground state. 
The network coordination number, which is the average number of bonds coming out of a node, is  denoted by $Z$. In order for such a network to be rigid, it needs to have an average $Z\geq Z_c$, the critical coordination number. In $d$ dimensions, and excluding finite-size effects, $Z_c = 2d$; the $2D$ networks used here have $Z>Z_c = 4$.

\begin{figure*}
\includegraphics[scale = 1]{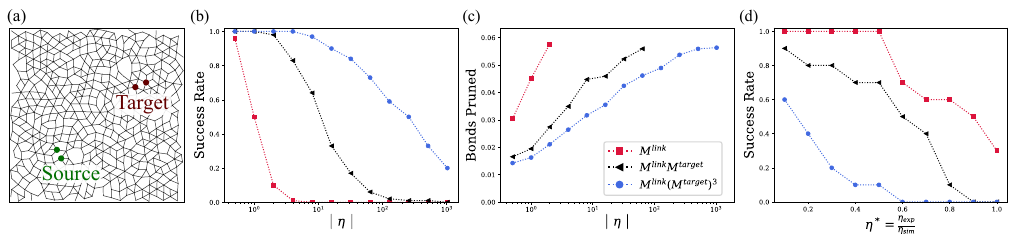}
\caption{(a) A sample network with adjacent nodes chosen to be source and target sites. (b) Results from pruning in simulations. Success rate of pruning networks as a function of $|\eta|$. Networks pruned to lower the effective moduli $M^{eff} = M^{link}(M^{target})^3$ (blue circles) have the highest success rate, followed by $M^{link}M^{target}$ (black triangles), with $M^{link}$ (red squares) being the least effective way to prune. (c) Average fraction of bonds pruned as a function of $|\eta|$ for each of the three $M^{eff}$ (d) Performance of experimental networks that were designed in simulations to have $\eta \approx 1$. Plot shows the fraction of networks with a response higher than $\eta^*$ as a function of $\eta^*$.}
\label{fig:simulations}
\end{figure*}

In order to incorporate a long distance `allosteric' response between two distant sites within a network, two pairs of nodes are picked at random as the source and target respectively. These are separated by typically half of the system size. One such network is shown in Fig.~\ref{fig:simulations}(a). In order to have an allosteric interaction, there should be an output strain, $\epsilon_T$, at the target pair of nodes when an input strain, $\epsilon_S$, is applied between the two source nodes. The ratio of output to input strains is $\eta \equiv \frac{\epsilon_T}{\epsilon_S}$. The aim is to incorporate an allosteric response, with a desired value of $\eta$ in the network by removing specific bonds from the network using a local pruning rule that uses only information that is available prior to the pruning itself.   

The general idea is to apply strains at both the input and output sites (in some cases simultaneously and sometimes separately) to discover which bonds should be removed in order to produce a stress at the target when the source is activated.  One might be tempted simply to minimize the energy for a specific mechanical behavior.  That is, one might consider removing the bonds that are under highest stress when both the source and target are simultaneously put under the desired strains.  This, however, would nearly always fail because the dominant energy for the strained system is often just due to the strains of the source and target irrespective of whether the source and target are applied simultaneously.  In case of designing an allosteric response, the goal is not merely to lower the energy for the input and output strains, but to create an interaction between the source and target sites. Thus, the source and target sites must communicate with each other. In order to achieve this, one needs to identify specifically the bonds that facilitate and the ones that hinder this allosteric response. By identifying and pruning the right set of bonds, it is possible to minimize the \textit{interaction energy} (not just the total energy of distortion) of the input and output strains.

We apply a deformation to our system, $\epsilon^k$. This $\epsilon^k$ could be a single strain applied between two points in the system or a combination of strains applied at various locations. Due to this applied strain $\epsilon^k$, each bond in the system experiences some stress. The stress in bond $j$ that appears due to $\epsilon^k$ is $S_j^k$. 
For example, $S_j^{source}$ is defined as the stress in bond $j$ as a result of a the input strain applied at the $source$.

Since all the calculations below are in the linear-response regime,
\[S_j^{-k} = -S_j^k;~~ S_j^{k+l} = S_j^k + S_j^l\]
One can calculate the energies in all the bonds of the network under any applied deformation: The energy, $U_j^k$, in bond $j$ when it is under a stress $S_j^k$ is:
\begin{equation}
U_j^k= \frac{1}{2}S_j^k \gamma_j^k
\end{equation} 
where $\gamma_j^k$ is the strain of the bond $j$ under applied external strain $\epsilon^k$. 
Since the total energy of the network is simply the sum of the energies of all the individual springs, the total energy stored in the network under an applied strain $\epsilon^k$ is $U^k = \sum_j U^k_j$. 

The modulus $M^k$ for any given deformation $\epsilon^k$ is defined as $U^k = \frac{1}{2}M^k (\epsilon^k)^2$.  It can be decomposed into the contributions of each bond: $M^k = \sum_j M^k_j$. 
$M_j^k$ is related to $S_j^k$ as follows:
\begin{equation}
U^k_j = \frac{1}{2}M_j^k (\epsilon^k)^2 =\frac{1}{2}S_j^k \gamma_j^k
\end{equation}
For a linear spring, $S_j$ and $\gamma_j$ only differ by a factor of the spring constant. This gives the following:
\begin{equation}
M_j^k \propto (S^k_j)^2
\label{Mdef}
\end{equation}
$M^k_j$ is the contribution of bond $j$ to $M^k$, where $M^k$ is the modulus for the deformation $\epsilon^k$. In section~\ref{setup} it will be shown that $M_j^k$ is a quantity that can be measured experimentally. Moreover, it is easier to measure $M_j^k$ than to measure $S_j^k$ or $\gamma_j^k$. Since the goal is to be able to prune the networks in experiments, the pruning protocols presented below will be based on measurements of $M_j^k$.

\subsection*{Pruning algorithm}
\label{pruning algorithm}
In order to incorporate an allosteric response, it is not particularly important how much energy is required to move the $source$ nodes apart as long as it results in an effect of the correct sign and magnitude at the $target$ site. Therefore, it is of little importance what are the individual values of ($S_j^{source}$) and ($S_j^{target}$); the important quantity is the product of the two, ($S_j^{source}$)($S_j^{target}$). This term identifies which bonds are most relevant to both $source$ and $target$ and pruning the bonds with the largest $(S_j^{source}S_j^{target})$ helps create an interaction between the $source$ and $target$ sites.

Consider applying the input strain at $source$ and output strain at $target$. This is represented as $S^{s+t}$.

\begin{equation}
S_j^{s+t} = S_j^{source}+S_j^{target}.   
\end{equation}
Similarly, applying the input strain at $source$ and the negative of output strain at $target$ is represented by $S^{s-t}$.

\begin{equation}
S_j^{s-t} = S_j^{source}-S_j^{target}.
\end{equation}{}

The moduli $M_j$ can be expressed in terms of $S_j$:

\begin{equation}
M_j^{s+t} \propto (S_j^{s+t})^2 =  (S_j^{source}+S_j^{target})^2  
\end{equation}{}
\begin{equation}
M_j^{s-t} \propto  (S_j^{source}-S_j^{target})^2  
\end{equation}{}
and
\begin{equation}
M_j^{link} \equiv M_j^{s+t}-M_j^{s-t} = 4(S_j^{source})(S_j^{target}).
\end{equation}{}

Note that taking the difference of the in-phase and out-of-phase terms produces the product of stresses due to applied strain at source and target sites. This term, $M^{link}$, links the effects of strains at both source and target and can be either positive or negative. One pruning protocol that would create an allosteric interaction between the source and target would be to prune those bonds in the network that have a maximum value of $M^{link}$.  

Because $M_j^{link} = (S_j^{source})(S_j^{target})$ is symmetric between source and target, the source and the target have been treated on an equal footing. If this were the only criterion for pruning bonds, then the effect on the target by activating the source would be the same as the effect on the source by activating the target. Thus, such a criterion would produce $\eta \approx 1$.  

However in many situations it might be preferable to have $\eta \ne 1$. For example one might want to create a strain at the target that is twice as large as the strain at the source (\textit{i.e.}, $\eta = 2$).  This would require that the symmetry between source and target be broken so that, for example, the target nodes are easier to move than the source nodes. One effective way to break the symmetry,is to bias the modulus by giving more weight to $S_j^{target}$ than to $S_j^{source}$. One way to do this is to prune the bond with maximum value of $(M_j^{link})(M_j^{target})^n$ where $n>0$. 

These combinations of moduli are referred to as the effective modulus, $M^{eff}$.  In the results presented in the next section below there are three examples:
\begin{enumerate}
    \item $(M^{eff,0}) = M^{link}$,
    \item $(M^{eff,1}) = M^{link}M^{target}$,
    \item $(M^{eff,3}) = M^{link}(M^{target})^3$.
\end{enumerate}

%????SUBSCRIPT J REFERS TO BOND J, need a different way to differentiate between various $M{eff}$???

Our pruning algorithm is as follows: 

\begin{enumerate}
    \item Calculate $M_j^{eff}$ for each bond $j$ in the network;
    \item Remove the bond with the maximum value of $M_j^{eff}$;
    \item Calculate the new value of $\eta$;
    \item Repeat until the desired value of $\eta$ is obtained. 
\end{enumerate}

By using effective moduli as the underlying quantity that controls a network's behavior, it is possible to incorporate responses in disordered networks using local rules alone. The rest of this paper explores the efficacy of this pruning approach using spring network simulations followed by an experimental method to measure $M^{eff}$ in order to incorporate allosteric responses \textit{in-situ} in physical systems.

\section{Simulation results}

In order to check the efficacy of these algorithms, I simulate the response of networks as the protocols are applied. The simulations can be performed on networks with periodic boundaries as well as ones with free boundaries. A free boundary network is created by cropping out a circular section from a periodic network. This often produces dangling bonds or zero modes which are eliminated by removing the relevant bonds and nodes from the edges of the cropped network. Since open boundary networks are easier to build in experiments, I use these networks to compare the response between simulations and experiments.

The simulation results shown here are performed on networks that have periodic boundaries with $\sim500$ particles and $\sim1080$ bonds. Unpruned networks have $\eta \approx 0.0$ on average between randomly chosen source and target sites. Networks are pruned until the desired value of $\eta$ is reached or until the process fails due to the creation of a zero mode in the system. $50$ different networks are pruned for both positive and negative values of $\eta$.  The networks used in these simulations have an average $\Delta Z = Z-Z_c \approx 0.32$. This corresponds to an excess of $\sim7\%$ bonds more than necessary to maintain rigidity.

The success rate of each of the pruning methods is shown in Fig.~\ref{fig:simulations}(b). As one might expect, if we prune for higher $|\eta|$, the success rate decreases. It is clear from this data that using just $M^{link}$ to prune a network is not the best strategy because, due to the symmetry between source and target, one can prune only to a maximum of $|\eta| \approx 1$. Even this response can be achieved only about half the time.

Biasing $M^{eff}$ towards the target improves the effectiveness of pruning significantly; using $M^{eff,3} = (M^{link})(M^{target})^3$ makes it possible to reach $|\eta| > 1000$. However, it is important to note that these calculations are performed in the linear-response regime of the network. Such a very large $|\eta|$ implies that the source strain must be extremely small in order for the target strain to be $1000$ times the source strain and still be in the linear regime. This makes it highly impractical to measure such a response in a laboratory system.

Fig.~\ref{fig:simulations}(c) shows the average fraction of bonds that need to be pruned as a function of $\eta$.  We see that $100\%$ of the networks fail after an average of $\sim 6\%$ bonds are removed. This is when nearly all the excess bonds above the rigidity threshold have been removed; therefore removing any subsequent bond has a high probability of creating a zero-energy mode.

\subsection*{Efficiency in Experiments}

\label{lasercut-simulations}

It is known from previous studies that linear spring simulations do not capture all the material details of a real network~\cite{reid2018auxetic,reid2019auxetic}. There are other interactions, such as angle bending forces, that are present in a real material. In order to test how well our pruning algorithms translate to real networks, we design 2D networks with open boundaries using the three mentioned protocols and then fabricate them in experiments. 

We took networks with free boundaries ranging between 110 and 150 particles in size and pruned each of them using our three protocols. We stopped pruning either once the network has achieved $\eta>1$ or once a zero mode is produced so that the pruning process failed. Since not all of our algorithms have a $100\%$ success rate, we chose 10 networks that could be pruned successfully using the three effective moduli, $(M^{eff,0})$, $(M^{eff,1})$, and $(M^{eff,3})$.  For consistency, in all three cases the same set of source and target nodes are used.  For any given starting network, each protocol removes a different set of bonds. I then laser cut 30 realizations of these networks (10 networks $\times$ 3 algorithms) and measured their responses in experiments. 

Our networks were lasercut from $1.5mm$ thick sheets of silicone rubber with a hardness of shore A70. The bonds were made thinner near the nodes to minimize angle-bending interactions in the networks as was done in previous work~\cite{rocks2017designing}. The ratio of the width of a bond to its average length is 1:6 with the bonds being half as wide near the nodes.  

In order to measure the observed $\eta$ of these lasercut networks, an input strain of $5\%$ is applied at the source and the output strain is measured at the target site. Figure~\ref{fig:simulations}(d) shows the response of these networks. Since each designed network has a slightly different value of $\eta$, we normalize our experimental results, $\eta_{exp}$ by the value of $\eta$ produced in the simulation, $\eta_{sim}$: $\eta^* = {\eta_{exp}}/{\eta_{sim}}$ and plot it along the abscissa. The ordinate shows the fraction of networks whose response exceeds a given $\eta^*$. If our simulations were a perfect model for the experimental systems, then the data in Fig.~\ref{fig:simulations}(d) would have been a horizontal line at $1.0$.

This data is surprising because some algorithms have much better agreement between experiments and simulations than others. Interestingly, Fig.~\ref{fig:simulations}(b) shows that pruning with $M^{eff,0} = M^{link}$ works only about $50\%$ of the time but when these networks are translated to experiments, they have a very high success rate. On the other hand, $M^{eff,3} =M^{link}(M^{target})^3$  works very well in simulations but not in experiments. This suggests that the disparity between experiments and simulations increases as the complexity of the pruning algorithm increases. I hypothesize that the inclusion of $(M^{target})^n$, increases the effect of the non-linear terms so that the predictions from simulation are farther from our experimental results.

\section{Pruning \textit{in-situ}}
\label{setup}

Our results show that linear spring models do not work perfectly for designing real materials. In order to see what is going on in the laboratory material, we need a way to measure the stresses in a physical network. This section presents experiments to measure the stress distribution in physical networks and use that information to prune the networks \textit{in-situ}.

\subsection*{Setup}

\begin{figure}
\includegraphics[scale=1]{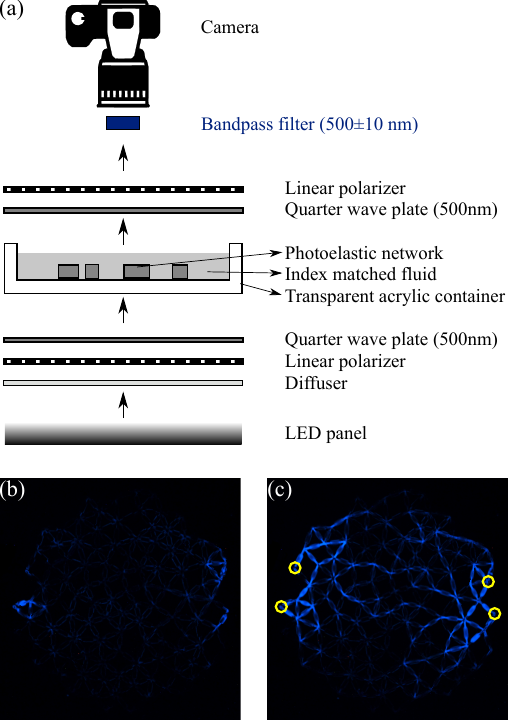}
\caption{(a) Setup for visualizing stresses in photoelastic networks (b) Sample image when the network has no external strain (c) Image of the same network as in (b) with applied strains between adjacent node pairs circled in yellow.}
\label{fig:setup}
\end{figure}

The stresses in a transparent material can be quantitatively detected by measuring stress-induced birefringence~\cite{hecht2002optics}.  The linear polarization of a beam of light will not be affected as it passes through an isotropic material; an analyzing polarizer with perpendicular orientation on the exiting side of the material will block all the light.  A photographic image would be completely dark.  However, if there are stresses in the photoelastic material, the polarization axis of the light will be rotated depending on the orientation of the stress with respect to the polarization axis.  The relative phase shift, $\Delta$ between two principal directions is proportional to the difference in the two principal stresses~\cite{hecht2002optics}. 
\begin{equation}
    \Delta \propto S_1 - S_2
\end{equation}
If the stresses are small enough so that the rotation angle is small, the analyzing polarizer will transmit the light in proportion to the stress. A photographic image will be bright in those regions where the stress is large and completely dark where there is no net stress.

Using circularly polarized light allows the magnitude of the stress to be measured regardless of its orientation.  The drawback of using circular polarization is that circular polarizers are sensitive to the wavelength of the light.  Thus monochromatic light must be used. 

In the experimental measurements of stresses reported here, the disordered networks are made out of molded urethane rubber (Smooth-on Clear Flex\texttrademark~ 50) with a shore hardness of A50. This material is highly sensitive to stresses and allows measurement of stresses in the range of $4\si{\kilo\pascal} - 20 \si{\kilo\pascal}$ in the current setup.  %???CAN YOU MAKE THIS MORE QUANTITATIVE.????  
The liquid urethane material is poured into molds in the shape of the desired networks.  The molds are 3D printed in a soft rubber material with a shore hardness of A28. Before each use, the molds were coated with a release agent (EaseRelease\texttrademark~200) to ensure that the molded urethane networks are easy to remove from the molds. The networks were cured in the mold at room temperature for a duration of 12 hours. After removal from the mold, they were cured for an additional 2-3 days at room temperature to reduce the tackiness of the surfaces of the molded networks.

A schematic of the experimental setup is shown in Fig.~\ref{fig:setup}(a). Data is collected by the camera at wavelength $\lambda = 500\pm10nm$.  The initially unpolarized white light is first polarized using a linear polarizer and then converted to circular polarization using a quarter-wave plate for $\lambda =500nm$. The light then passes through the sample.  On the other side of the sample, another quarter-wave plate  converts the light back to linearly polarized light followed by a linear polarizer that is oriented perpendicular to the polarization of the initial light. 
The total stress at each point in the material can be measured by the intensity of the light in the photographic image. %Since the pruning protocols described above are only sensitive to the magnitude of stresses, using circular polarizers makes the measurements easier.

Even after letting the molded photoelastic networks cure for a few days, their surfaces are still tacky enough to stick to a glass or acrylic surface.  This produces extraneous stresses in the material that are unrelated to those caused simply by placing strains at the source or target nodes. Additionally, because of a high surface tension of the molding liquid, the top surface of the network has a meniscus so that it is not completely flat. As a result, a ray of light going through such a curved surface is reflected and refracted in various directions and gives rise to unwanted signals in our data. 

Both of these problems can be eliminated by submerging the networks in an index-matched fluid. I used mineral oil with a refractive index of $1.47$, which is very close to the refractive index of the photoelastic networks which is $1.48$.
Sample images from this setup with and without an applied strain are shown in Fig.~\ref{fig:setup}(b) and (c) respectively.
The image in Fig.~\ref{fig:setup}(b) shows that when the network is under no stress, the image of the network in the camera is dark and nearly undetectable. 
%When the photoelastic network is under stress, it changes the polarization of light passing through it. The stress optic law states that the relative phase shift, $\Delta$ between two principal directions is proportional to the difference in the two principal stresses\cite{hecht2002optics}.
When an external strain is applied to the network as shown in Fig.~\ref{fig:setup}(c), different bonds in the network are stressed by different amounts as a reaction to the input strain. These stressed regions now appear as bright spots in our image.

The intensity of light, $I(x,y)$ at any point in the material $(x,y)$ is related to the electric field, $E(x,y)$, and the phase shift, $\Delta (x,y)$:
\begin{equation}
    I(x,y) \propto E(x,y)^2 \propto \sin(\Delta(x,y))^2 \propto \sin(S(x,y))^2.
\end{equation}
As long as the input strain is sufficiently small, the resultant stresses are in the regime where $\sin(S) \approx S$. In this limit, the brightness of a bond is proportional to the square of stress. Hence, the intensity $I$ averaged over the length of bond $j$ is proportional to $M_j$ in Eq.~\ref{Mdef}.

\subsection*{Experimental Results}

\begin{figure}
\includegraphics[scale = 1]{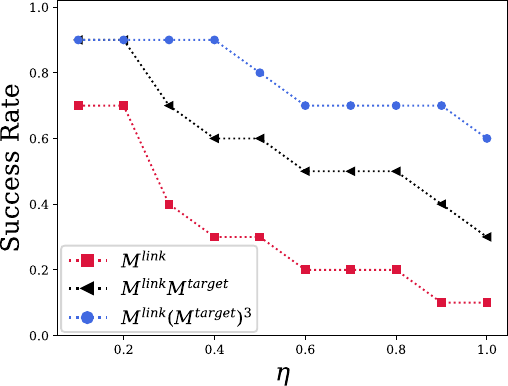}
\caption{Success rate of networks pruned \textit{in-situ} as a function of $\eta$. Following the same trend as the simulations, networks pruned to lower the effective moduli $M^{eff} = M^{link}(M^{target})^3$ (blue circles) have the highest success rate, followed by $M^{eff} =M^{link}M^{target}$ (black triangles), and $M^{eff} =M^{link}$ (red squares). }
\label{fig:photoelastic_data}
\end{figure}

I conduct the \textit{in-situ} pruning experiments on photoelastic networks that are between $110$ and $150$ nodes in size. The photoelastic networks are molded to be 5mm thick with an average bond length of 12mm. The ratio of the width of a bond to its average length is $1:4$ with the bonds being a third as wide near the nodes than in their middle. This slight difference in geometry makes the middle part of the bonds wider than in the experiments described above that pruned bonds according to the algorithm based on simulation results.
It ensures that the effect of  noise from the edges of photoelastic bonds is minimized. 
In order to measure the response of these pruned networks, I remake them in silicone rubber using the same specifications as in section~\ref{lasercut-simulations}. This is done to avoid any differences arising from the material properties while comparing the response of designed and experimentally pruned networks.

In order to prune these networks, I take images where a strain is applied (i) at the source and target in phase $(M^{s+t})$, (ii) out of phase $(M^{s-t})$, and (iii) only at the target site $(M^{target})$. By computing combinations of these images, I obtain a measure of the same three effective moduli as defined in section~\ref{pruning algorithm}: $M^{eff} = M^{link}$, $M^{eff} =M^{link}M^{target}$, and $M^{eff} =M^{link}(M^{target})^3$. The experiments were conducted on $10$ different networks for each $M^{eff}$. In each pruning step, I first calculate $M^{eff}$ for all the bonds, remove the bond with maximum $M^{eff}$, and then measure $\eta$ of the pruned network. As the bonds are pruned sequentially, the signal-to-noise ratio drops steadily; when the signal is too low to be reliable, the pruning is halted. The recorded $\eta$ for a particular sample is the maximum $\eta$ that the network exhibits at any point during the pruning process. 

The results from \textit{in-situ} pruning experiments are shown in Fig.~\ref{fig:photoelastic_data}. To some extent, all pruning methods show an allosteric response. In particular, by sequentially removing the bond with the largest value of $M^{eff,3} = M^{link}(M^{target})^3$ at each step, networks can often be successfully pruned to produce $\eta = 1$. The three different pruning methods have very different success rates.  This was also seen in the simulation results shown in Fig.~\ref{fig:simulations}(b). The results from the experiments and from the simulations follow the same trend. %and confirm that our pruning algorithms in experiments work as we expected. 

The same network design pruned to minimize the same $M^{eff}$ leads to the removal of very different sets of bonds in simulations versus those pruned in the photoelastic experiments. This is not surprising since the simulations only considered central, harmonic springs, whereas the experiments had all the interactions inherent in an elastic sheet such as angle-bending forces at each node and non-linear stress/strain curves for all the bonds. 
However, it is surprising that, compared to the experimental results shown in Fig.~\ref{fig:simulations}(d), the trend for networks pruned \textit{in-situ} is completely reversed. 
This suggests that the simulation models are too simple to capture all the stresses in a real material.

These results reinforce the conclusion that in order to make a physical system with an allosteric response, a simplified model is not sufficient and it is important to visualize and measure all the interactions in the system. Using an experimental procedure to detect the stresses in physical networks, it is possible to create allosteric networks in experiments with a high success rate.

\section{Conclusions}

The emphasis of this work has been to understand the local pruning rules that control allosteric response in mechanical networks as well as to fabricate these networks in real (physical) experimental systems. In order to create a robust response, it is important to identify the relevant parameters that govern allosteric behavior.
The effective moduli introduced here are the bond-level contributions to allostery. The simulations show that this approach is very effective at designing allosteric responses with minimal computational cost. 
In some cases, the results of these simulations can be directly translated into experimental realizations. These local pruning rules are more efficient than previous methods for incorporating allosteric behavior into experimental systems~\cite{rocks2017designing}. 
Similar local pruning rules can also be used to create more general responses such as multiple pairs of allosteric source and target sites, or a single source site controlling multiple target sites. As expected, when the complexity of the incorporated response increases, the efficiency of such pruning algorithms goes down.

As expected, however, the simulations of networks connected by linear, central-force springs do not capture all the details of a real material. In order to  create allosteric responses efficiently in experiments, it is necessary to be able to measure stress distributions in the physical networks. I have presented an experimental procedure to visualize and measure the stresses in these systems and use them to prune the appropriate bonds in the networks \textit{in-situ} to achieve a desired response.

The experimental technique to measure local stresses provides a pathway for modifying physical systems with more complex interactions. It would be interesting to see if it can be used to create other mechanical responses, such as auxetic behavior. This protocol can also be applied to other disordered systems that are not based on spring networks. The pruning methods explored here could be extended to 3D systems and to smaller length scales by using stimuli responsive materials that can detect stresses in polymers at the molecular level~\cite{chen2020force,deneke2020engineer}. 

Recent work has shown that stress-induced aging can be used to modify material properties~\cite{pashine2019directed}. An externally imposed strain can \textit{direct} the evolution of a material and determine its mechanical response. Our understanding of local pruning rules that control allostery in combination with directed aging protocols can be very useful in designing allosteric systems while eliminating the need to manipulate the material manually at the microscopic level. 
Local learning rules, such as the ones presented here, are also of interest in supervised learning of elastic and flow networks~\cite{stern2020supervised}

In conclusion, local pruning rules that allow the manipulation of material response combined with the ability to measure stress distributions \textit{in-situ}, enable the modification of materials that are more complex than the linear spring networks that have been the focus of previous simulation studies. This approach opens up novel ways to build mechanical metamaterials with a desired function without relying on computer simulations.

\section{Acknowledgements}

I would like to thank Daniel Hexner, Andrea J. Liu, and Nachi Stern for insightful discussions. I am deeply grateful to Sidney R. Nagel for his advising and mentoring. I would also like to thank Cacey Stevens Bester for information regarding photoelastic materials, and Robert Morton for help with 3D printing.  This work was supported by the NSF MRSEC Program DMR-2011854 (for experimental studies), the US Department of Energy, Office of Science, Basic Energy Sciences, under Grant DE-SC0020972 (for theoretical model development) and by the Simons Foundation for the collaboration ``Cracking the Glass Problem'' Award 348125 (for simulations).

\bibliographystyle{unsrt}
\bibliography{biblio}

\end{document}